\documentclass[conference]{IEEEtran}

\usepackage{amsmath,amssymb}
\usepackage{graphicx}
\usepackage{subcaption}
\usepackage{enumitem}
\usepackage[font=small,labelfont=bf]{caption}

\usepackage{booktabs,tabularx,threeparttable}
\usepackage{ragged2e}
\usepackage{pifont}
\usepackage{xcolor}
\usepackage{microtype}

\usepackage[table]{xcolor}   % for \rowcolor
\usepackage{tabularx}       % tabularx environment
\usepackage{booktabs}       % \toprule, \midrule, \bottomrule
\usepackage{threeparttable} % threeparttable environment
\usepackage{array}          % for custom column types

\newcolumntype{L}{>{\raggedright\arraybackslash}X}
\newcolumntype{C}[1]{>{\centering\arraybackslash}p{#1}}

\usepackage{needspace}
\usepackage{stfloats}

\usepackage[hidelinks]{hyperref}

\newcolumntype{J}{>{\justifying\arraybackslash\setlength{\parindent}{0pt}}X} % justified
\newcolumntype{I}{>{\itshape\justifying\arraybackslash\setlength{\parindent}{0pt}}X} % italic + justified
\newcolumntype{C}[1]{>{\centering\arraybackslash}p{#1}}

\newcommand{\cmark}{\textcolor{green!60!black}{\ding{51}}}
\newcommand{\xmark}{\textcolor{red!70!black}{\ding{55}}}
\newcommand{\meh}{\textcolor{yellow!60!black}{\large\textbullet}}

\title{\title{Toward Noise-Aware Audio Deepfake Detection: Survey, SNR-Benchmarks, and Practical Recipes}}        

\author{
    \IEEEauthorblockN{Udayon Sen\IEEEauthorrefmark{1},
    Alka Luqman\IEEEauthorrefmark{1}\IEEEauthorrefmark{2},
    Anupam Chattopadhyay\IEEEauthorrefmark{1}}
    \IEEEauthorblockA{\IEEEauthorrefmark{1}Nanyang Technological University, Singapore}
    \IEEEauthorblockA{\IEEEauthorrefmark{2}Deepfaic}
}

\begin{document}
\maketitle

% ===================== Abstract =====================
\begin{abstract}
Deepfake audio detection has progressed rapidly with strong pre-trained encoders (e.g., WavLM, Wav2Vec2, MMS). However, performance in \emph{realistic capture conditions}—background noise (domestic / office / transport), room reverberation, and consumer channels—often lags clean-lab results. We survey and evaluate robustness for state-of-the-art audio deepfake detection models and present a reproducible framework that mixes MS-SNSD noises with ASVspoof 2021 DF utterances to evaluate under \emph{controlled signal-to-noise ratios (SNRs)}. SNR is a measured proxy for noise severity used widely in speech; it lets us sweep from near-clean ($35$\,dB) to very noisy ($-5$\,dB) to quantify graceful degradation \cite{specaugment,park2020mtr,ms_snsd}. We study multi-condition training and fixed-SNR testing for pretrained encoders (WavLM, Wav2Vec2, MMS), reporting accuracy, ROC--AUC, and EER on binary and four-class (authenticity$\times$corruption) tasks. In our experiments, finetuning reduces EER by $\sim$10--15 percentage points at 10--0\,dB SNR across backbones.
\end{abstract}

% ===================== Introduction =====================
\section{Introduction}

Recent advances in generative models, especially in human-like speech synthesis, have made it possible to not only replicate human speech but make it almost indistinguishable from genuine recordings \cite{tts_survey_2023,voice_conversion_survey_2020,diffusion_tts}. 

These so-called \emph{audio deepfakes} are produced using various techniques, the most popular being text-to-speech (TTS), voice conversion (VC), and diffusion-based generative models \cite{tts_survey_2023,voice_conversion_survey_2020,diffusion_tts}. Unlike earlier systems like rule-based or concatenative audio synthesis, modern architectures are able to imitate speaker identity and characteristics such as prosody and tone in a convincing manner, which blurs the boundary between real and synthetic speech.

The applications for these techniques span accessibility and extend into domains like entertainment and music, but also into highly sensitive and risky scenarios such as voice biometrics for account access and high-stakes phone scams in finance and politics \cite{asvspoof2019,asvspoof2021,voice_biometrics_survey,deepfake_misinfo_survey}.

As a result, deepfake audio detection has become a critical problem and an important area of ongoing research to develop reliable, robust, and generalizable detection methods.

There has been a significant amount of research with numerous deepfake detection models and methods, especially benchmarked through the ASVspoof Challenge series \cite{asvspoof2019,asvspoof2021}. These benchmarks catalyzed progress by providing large-scale datasets, standardized protocols, and evaluation metrics such as the Equal Error Rate (EER) \cite{eerlit}. Earlier methods relied on specialized acoustic features like Linear Frequency Cepstral Coefficients (LFCC) and Constant-Q Cepstral Coefficients (CQCC), often with traditional classifiers \cite{lfcc,cqcc}. Newer systems leverage deep neural networks and self-supervised speech representations for better accuracy. 

\paragraph*{\textbf{Research gap}}
However, despite notable gains in controlled test conditions, however, these systems often suffer from poor generalization with data ``in-the-wild'' \cite{aasist,rawnet2,rawnet3}. \textit{Table~\ref{tab:related}} summarizes representative prior work and the extent to which noise robustness is addressed.

Another major problem is the assumption of clean laboratory conditions. In practice, real-world audio is rarely pristine: speech is captured through consumer microphones or telephones, or in noisy conditions like environmental noise, reverberation, and channel distortions, which corrupt the speech data itself. Such conditions degrade the performance of deepfake detectors and reduce reliability, leading to increased false positives (labeling genuine speech as spoof) or false negatives (failing to catch a deepfake). Thus, the robustness gap between research prototypes and effective real-world deployable systems remains significant.

\paragraph*{\textbf{Related work}} 
The problem of robustness under noise is not unique to deepfake detection alone but also arises in speech technology more broadly. In automatic speech recognition (ASR), a lot of research has been done on mitigating noise effects via signal enhancement, robust feature extraction, and multi-condition training \cite{specaugment,park2020mtr,ms_snsd}. Data augmentation technologies such as SpecAugment, length perturbation, and additive noise from corpora such as MS\textendash SNSD (Microsoft Scalable Noisy Speech Dataset) are now standard in ASR pipelines \cite{specaugment,park2020mtr,ms_snsd}. However, for audio deepfake detection, such practices have only been recently adopted and systematic evaluations of noise robustness are scarce \cite{wang2020noise,codec_aug_spoof,channel_aug_spoof,ssl_spoof_probe}.

The mismatch between training on clean corpora and deployment in noisy real-world environments creates a pressing need for controlled studies of noise-aware detection. In particular, foundation models such as Wav2Vec2 \cite{wav2vec2}, WavLM \cite{wavlm}, and MMS \cite{mms}, which have proved powerful in downstream ASR and speaker tasks, have not been comprehensively benchmarked under spoofing conditions across varying \emph{signal-to-noise ratios (SNRs)}. Understanding their limitations and strengths under noise corruption can guide researchers and practitioners towards developing more resilient detection systems.

In speech processing, SNR provides a task-agnostic measure of additive-noise severity: the lower the SNR, the more the noise. It is interpretable, controllable, and comparable across datasets; ASR and enhancement literature routinely reports per-SNR curves for the same reasons \cite{specaugment,park2020mtr,ms_snsd}. Using SNR therefore lets us isolate noise effects from other confounding factors (speakers, content, codec) and benchmark robustness consistently.

The role of SNR in robustness evaluation is well established in classical ASR research. Early benchmarks such as Aurora-4 and the CHiME challenges systematically tested recognition systems under controlled SNRs and noisy conditions. Results have consistently shown a monotonic degradation of performance as SNR decreases, with multi-condition training and noise-aware features improving robustness. By drawing on these practices, our framework aims to adopt a familiar and interpretable setup for the deepfake detection task, enabling consistency with the speech-robustness literature.

\paragraph*{\textbf{Our contribution}} 

This paper aims to address the gap by providing both a survey and an empirical benchmark of noise robustness in audio deepfake detection models. Our contributions are threefold:
\begin{enumerate}[label=\roman*.]
    \item We review prior work on spoofing detection and robustness, situating the problem within broader speech processing research and highlighting open challenges.
    \item We introduce a reproducible experimental framework that integrates the ASVspoof 2021 corpus with noise augmentation across varying degrees from the Microsoft Scalable Noisy Speech Dataset (MS\textendash SNSD), which contains various forms of ambient noise, enabling systematic evaluation under controlled SNR conditions.
    \item We conduct empirical studies of state-of-the-art pretrained speech encoders (Wav2Vec2, WavLM, MMS) in both binary (real vs.\ spoof) and multi-class (real/spoof $\times$ clean/noisy) detection setups. We compare frozen (baseline) versus finetuned configurations and analyze how performance degrades as noise severity increases.
\end{enumerate}

% ===================== Related Work Table (Table 1) =====================
\begin{table*}[!t]
\centering
{\small
\setlength{\tabcolsep}{6pt}
\renewcommand{\arraystretch}{1.15}
\begin{threeparttable}
\caption{Summary of related work on audio deepfake detection and noise robustness.
\emph{Noise legend:} \cmark{} = robust (desirable), \meh{} = limited (moderate), \xmark{} = not robust (undesirable).}
\label{tab:related}
% columns: Work (flexible), Dataset(s) (fixed width), Noise (fixed width centered), Key Contribution (flexible)
\begin{tabularx}{\textwidth}{@{} L C{3.2cm} C{1.6cm} X @{}}
\toprule
\textbf{\upshape Work} & \textbf{Dataset(s)} & \textbf{Noise} & \textbf{Key Contribution} \\
\midrule
ASVspoof 2019~\cite{asvspoof2019} & ASVspoof 2019 (LA, PA) & \xmark &
Established benchmarks for logical/physical access attacks. \\[4pt]
ASVspoof 2021~\cite{asvspoof2021} & ASVspoof 2021 (DF, LA, PA) & \meh &
Introduced deepfake speech corpus; considered replay noise scenarios. \\[4pt]
RawNet2~\cite{rawnet2} & ASVspoof 2019 & \xmark &
End-to-end raw waveform model for spoof detection. \\[4pt]
AASIST~\cite{aasist} & ASVspoof 2021 & \xmark &
Graph-based spectral\textendash temporal attention for deepfake detection. \\[4pt]
Noise-Robust Spoofing~\cite{wang2020noise} & ASVspoof 2019 (+ noisy variants) & \cmark &
Evaluated additive-noise augmentation and multi-condition training. \\[4pt]
Wav2Vec2 Detectors~\cite{wav2vec2_spoof} & ASVspoof 2021 & \xmark &
Leveraged pretrained self-supervised encoders for detection. \\[6pt]
\rowcolor{gray!15}
\textbf{This work} & \textbf{ASVspoof 2021 + MS\textendash SNSD} & \textbf{\cmark} &
\textbf{Unified benchmarking of pretrained encoders under noise for binary and 4-class tasks.} \\
\bottomrule
\end{tabularx}
\begin{tablenotes}[flushleft]
\footnotesize
\item \textbf{Abbreviations:} DF = Deepfake, LA = Logical Access, PA = Physical Access, MS\textendash SNSD = Microsoft Scalable Noisy Speech Dataset.
\end{tablenotes}
\end{threeparttable}
}
\end{table*}

\begin{figure*}[!t]
  \centering
  \includegraphics[width=0.9\textwidth]{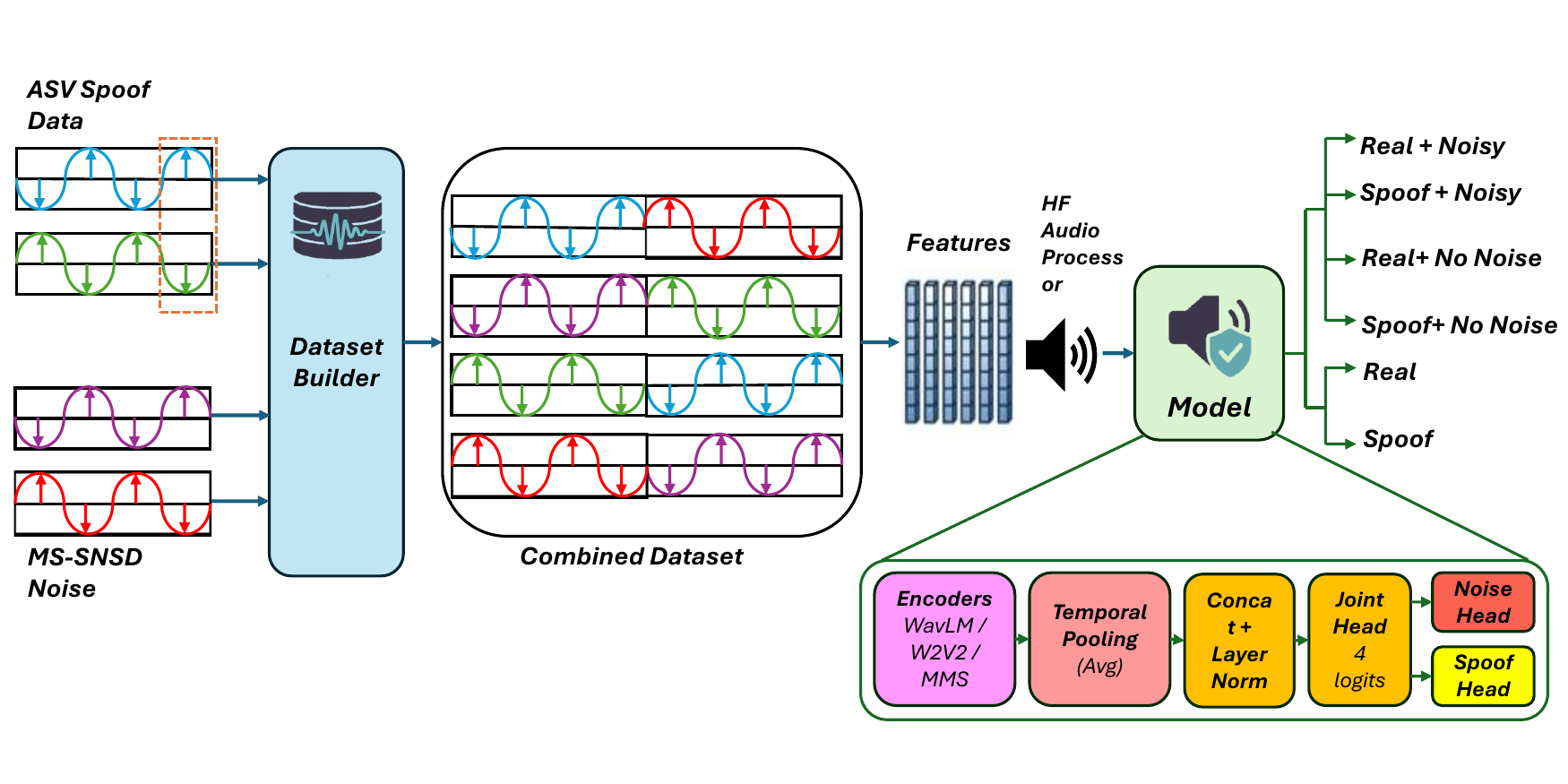}
  \caption{Dataset formation and evaluation design: ASVspoof 2021 (DF) speech mixed on-the-fly with MS-SNSD noises; mixed and fixed-SNR test conditions.}
  \label{fig:pipeline}
\end{figure*}

To make these evaluations concrete, Figure~\ref{fig:pipeline} outlines the dataset formation and evaluation design we use throughout the paper.

\section{Methodology and Experimental Design}

We construct a controlled and reproducible experimental framework to examine deepfake-speech detection under varied acoustic conditions. Two complementary resources are used: (i) \textbf{ASVspoof 2021 (DF)}, providing standardized protocols for bonafide and spoofed speech, and (ii) \textbf{MS-SNSD}, offering diverse ambient noise categories (domestic, office, outdoor, transport). ASVspoof serves as the base speech corpus, while MS-SNSD supplies additive noise signals at configurable SNRs, enabling systematic noise robustness without replay-channel confounds. 
\subsection{Task Formulations}

We explore two supervised classification settings:

\begin{itemize}
    \item \textbf{Binary Classification}: Real vs.\ Spoof.  
    This follows the commonly appropriate deepfake-detection setup and supports direct comparability with prior work. Evaluation uses accuracy, ROC--AUC, and equal error rate (EER).
    
    \item \textbf{Four-Class Classification}: Real+Clean, Real+Noisy, Spoof+Clean, Spoof+Noisy.  
    This configuration is aimed to decouple authenticity and acoustic corruption by labeling both factors explicitly. It enables structured analysis of how additive noise interacts with the decision space and supports models employing auxiliary noise-estimation heads

\end{itemize}

\subsection{Noise Augmentation Protocol}

To emulate diverse acoustic conditions while maintaining reproducibility, we adopt a two-stage augmentation protocol:

\begin{itemize}

    \item \textbf{On-the-Fly Multi-Condition Training}.  
    With probability $p = 0.5$, a speech segment is mixed with a randomly selected MS-SNSD clip. The SNR is sampled from  
    \[
    \{35, 30, 25, 20, 15, 10, 5, 0, -5\}\,\text{dB}.
    \]
    Occasionally, two noise clips are summed to increase non-stationarity. This procedure which is stochastic in nature exposes the model to a wide range of acoustic conditions without requiring pre-generated augmented sets.

    \item \textbf{Fixed-SNR Evaluation}.  
    For controlled testing, we generate per-SNR test splits in which all samples are corrupted at a single SNR value selected from the same grid. This ensures consistent severity-level comparisons and yields SNR-conditioned performance curves.
\end{itemize}

Unless stated otherwise, the mixed test set contains \textbf{80\% noisy} trials with randomly sampled SNRs, while the development set contains \textbf{20\% noisy} trials.

\subsection{Encoders and Prediction Heads}

We benchmark three self-supervised speech encoders—\textbf{WavLM-base+}, \textbf{Wav2Vec2-base}, and \textbf{MMS-300M}—as well as \emph{pairwise fusion} variants (e.g., WavLM--Wav2Vec2, WavLM--MMS).  
Each encoder feeds into a lightweight classification head:
\[
\text{LayerNorm} \rightarrow \text{Dropout}(0.1) \rightarrow \text{Linear}.
\]

Two training modes are examined:
\begin{itemize}
    \item \textbf{Frozen}: The encoder is kept fixed and only the head parameters are updated.
    \item \textbf{Finetuned}: All encoder parameters are optimized end-to-end.
\end{itemize}

\begin{figure}[htbp]
  \centering
  \includegraphics[width=\columnwidth]{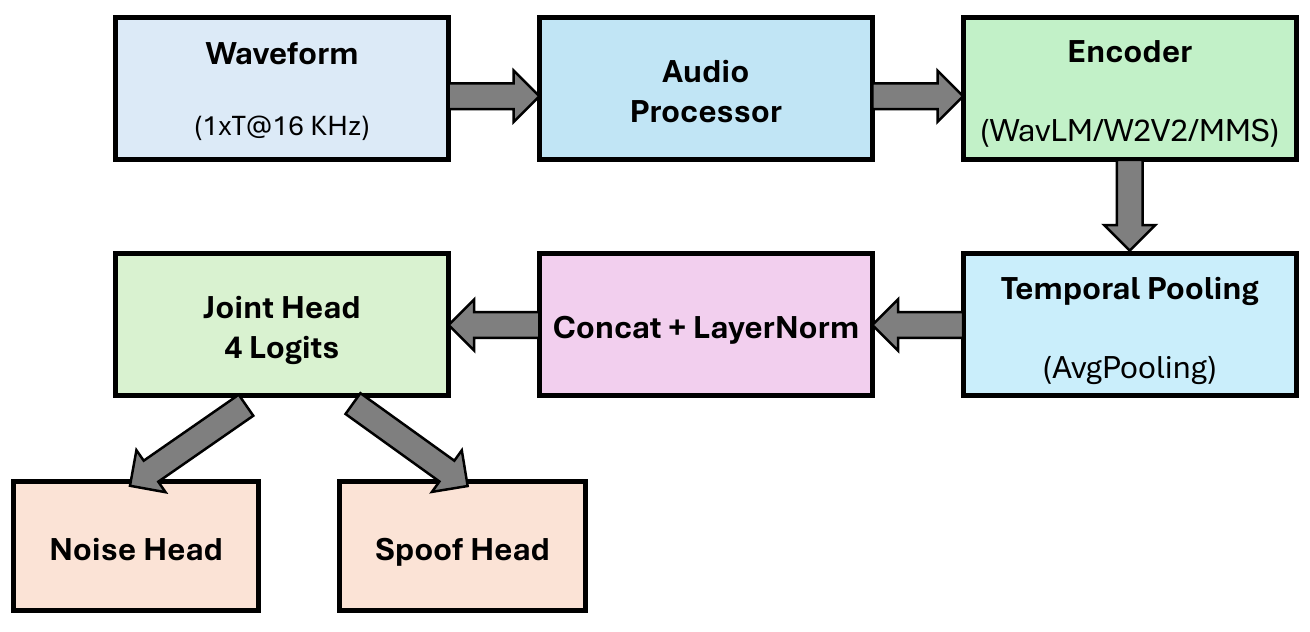}
  \caption{Model overview: waveform $\rightarrow$ feature extractor $\rightarrow$ SSL encoder(s) $\rightarrow$ temporal pooling $\rightarrow$ prediction head with spoof and optional noise logits.}
  \label{fig:model}
\end{figure}

The full architecture, including temporal pooling and the optional joint spoof/noise head, is illustrated in Fig.~\ref{fig:model}.

% ===================== Model Overview (Fig. 2) =====================

\subsection{Optimization and Training Schedule}

Optimization uses AdamW with weight decay $1\times10^{-2}$. Training proceeds in two phases:
\begin{itemize}
    \item \textbf{Head-only warmup}: learning rate $3\times10^{-4}$ for 3 epochs.
    \item \textbf{End-to-end finetuning}: learning rate $3\times10^{-5}$ for 5 epochs.
\end{itemize}

Mixed-precision training (\texttt{torch.amp}) is used.  
Speech segments are \textbf{2\,s} long by default, with an ablation on \textbf{4\,s}.  
Class imbalance is handled using a weighted sampler (binary) or class-weighted cross-entropy (four-class).  
Default batch sizes are 12 (train) and 16 (evaluation), determined by GPU memory constraints.

\subsection{Evaluation Metrics}

We report three complementary metrics—accuracy, ROC--AUC, and EER—that together capture threshold-free ranking quality, threshold-specific operating points, and overall correctness.

\paragraph*{Accuracy}
\textbf{Accuracy} reflects the proportion of correctly classified utterances under a fixed decision threshold. Although intuitive, it is sensitive to class imbalance and does not reveal how well models rank spoofed versus bonafide trials when score distributions overlap. We therefore treat accuracy as a coarse indicator and rely more heavily on threshold-free metrics for robustness analysis.

\paragraph*{ROC--AUC}
\textbf{ROC--AUC} (Area Under the Receiver Operating Characteristic Curve) quantifies the probability that a randomly sampled spoof trial receives a lower score than a randomly sampled bonafide trial. It is independent of the classification threshold and therefore robust to skewed class priors and application-specific operating ranges. In the context of our work—where noise shifts score distributions—ROC--AUC provides a stable ranking-based view of robustness across SNRs and backbones.

\paragraph*{Equal Error Rate (EER)}
\textbf{EER} denotes the point on the ROC curve at which the false acceptance rate equals the false rejection rate. EER is widely adopted in speaker verification and spoofing literature because it directly reflects the operating point where two types of errors trade off most symmetrically. In noisy conditions, EER rises sharply, making it a sensitive indicator of how corruption affects separability between real and spoofed speech. We therefore use EER as the primary robustness metric in our per-SNR analyses.

\medskip
Metrics are computed over all trials for each condition. Confidence intervals are omitted due to space constraints but can be incorporated in future benchmark versions for comparison.

% \subsection{Reproducibility (To be made available)}
% We release: (i) augmentation specification (noise directories, SNR grid, probabilities), (ii) data loaders for \emph{mixed} and \emph{fixed-SNR} evaluation, and (iii) a single script that runs the full sweep across backbones, including frozen and finetuned variants. 

% ===================== Mixed-test AUC pair (Fig. 3) =====================
\begin{figure*}[!t]
  \centering
  \begin{subfigure}[t]{0.49\textwidth}
    \centering
    \includegraphics[width=\linewidth]{\detokenize{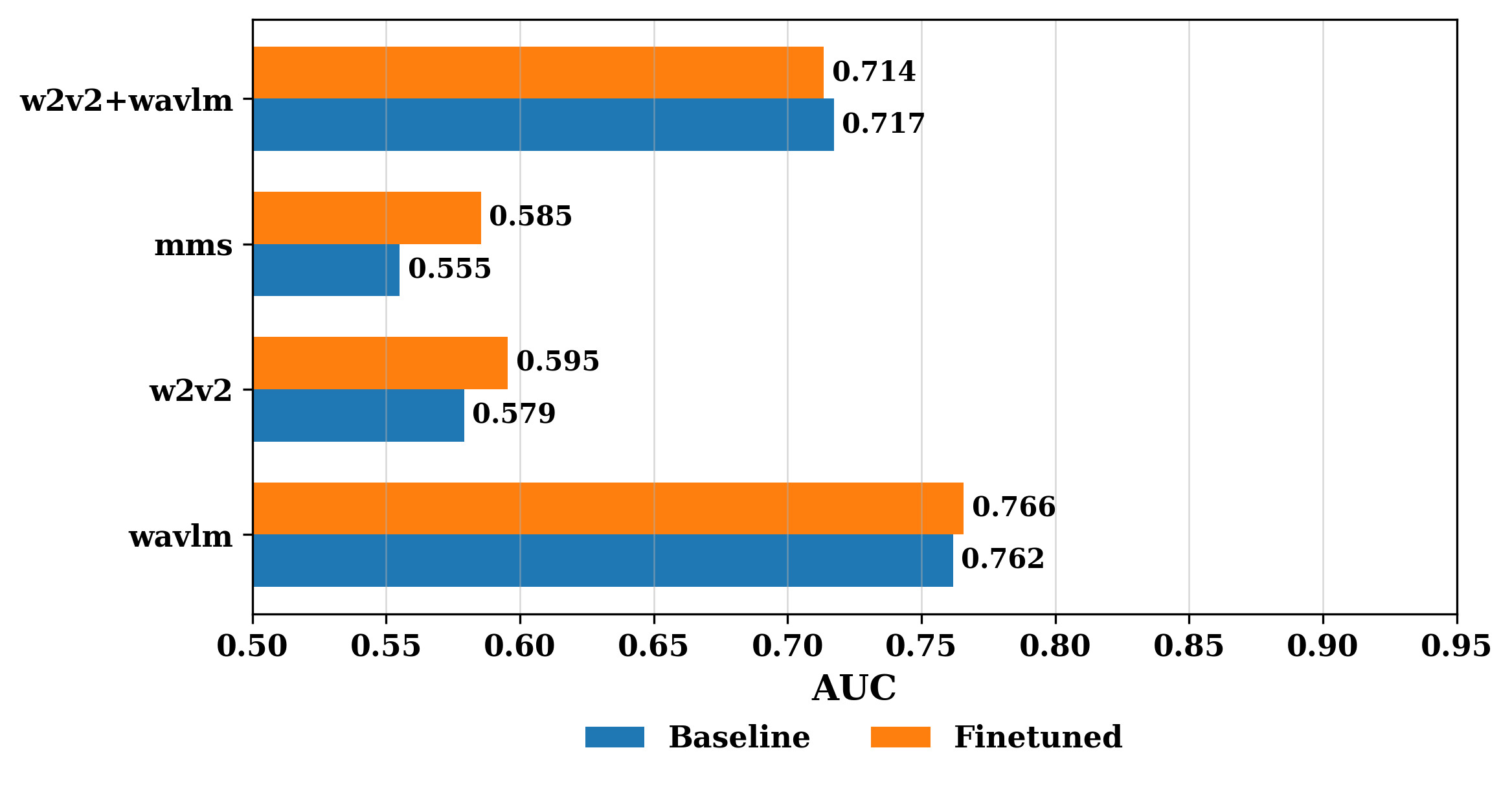}}
    \caption{Mixed-test ROC--AUC for \emph{spoof detection} (real vs.\ spoof). Bars compare frozen baselines to finetuned variants; higher is better.}
    \label{fig:mixed_spoof_auc}
  \end{subfigure}\hfill
  \begin{subfigure}[t]{0.49\textwidth}
    \centering
    \includegraphics[width=\linewidth]{\detokenize{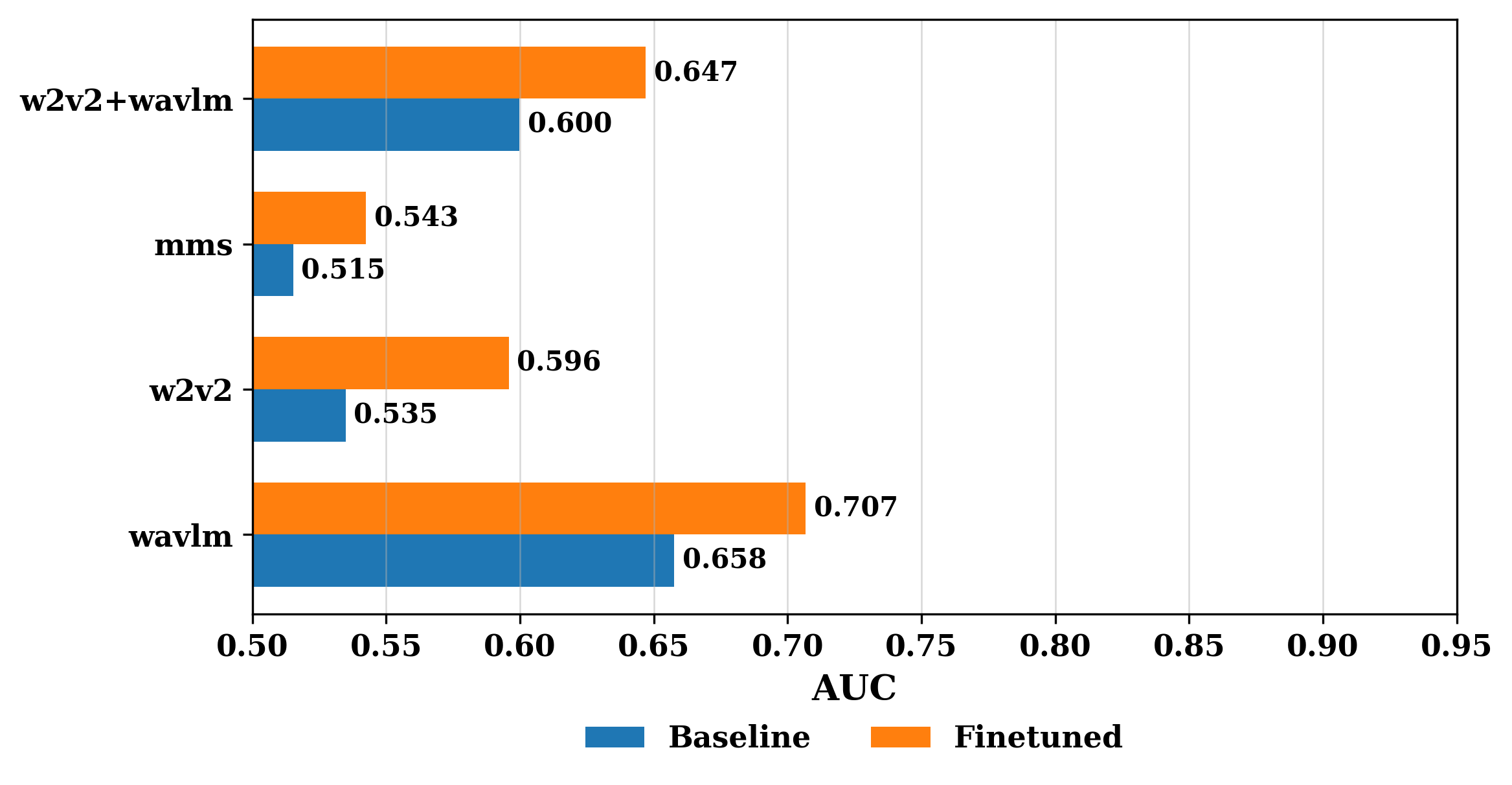}}
    \caption{Mixed-test ROC--AUC for \emph{noise discrimination} (clean vs.\ noisy). Bars compare frozen baselines to finetuned variants; higher is better.}
    \label{fig:mixed_noise_auc}
  \end{subfigure}
  \caption{Mixed-test ROC--AUC comparisons. Panel~(a) summarizes spoof detection; panel~(b) summarizes noise discrimination.}
  \label{fig:mixed_auc_combo}
\end{figure*}

% ===================== Results =====================
\section{Experimental Results}
\noindent We report results under mixed- and fixed-SNR conditions; detailed per-SNR analyses follow. Across backbones, finetuning yields the largest absolute EER gains (about 10--15 percentage points) at mid/low SNRs (10--0\,dB). 

\subsection{Overview: Mixed-Test ROC--AUC}
Across realistic SNR mixtures, finetuning consistently improves both spoof and noise discrimination for all backbones (Figure~\ref{fig:mixed_auc_combo}). The gains are modest in near-clean portions of the mix and widen as the proportion of mid/low-SNR trials increases, indicating better discounting of additive noise. WavLM typically attains the highest mixed-test AUCs, with Wav2Vec2 and MMS following closely depending on the noise composition. These aggregate trends mirror the fixed-SNR analyses reported next.

% ===================== Binary AUC (Fig. 4) =====================
\begin{figure}[t]
  \centering
  \includegraphics[width=0.8\linewidth]{\detokenize{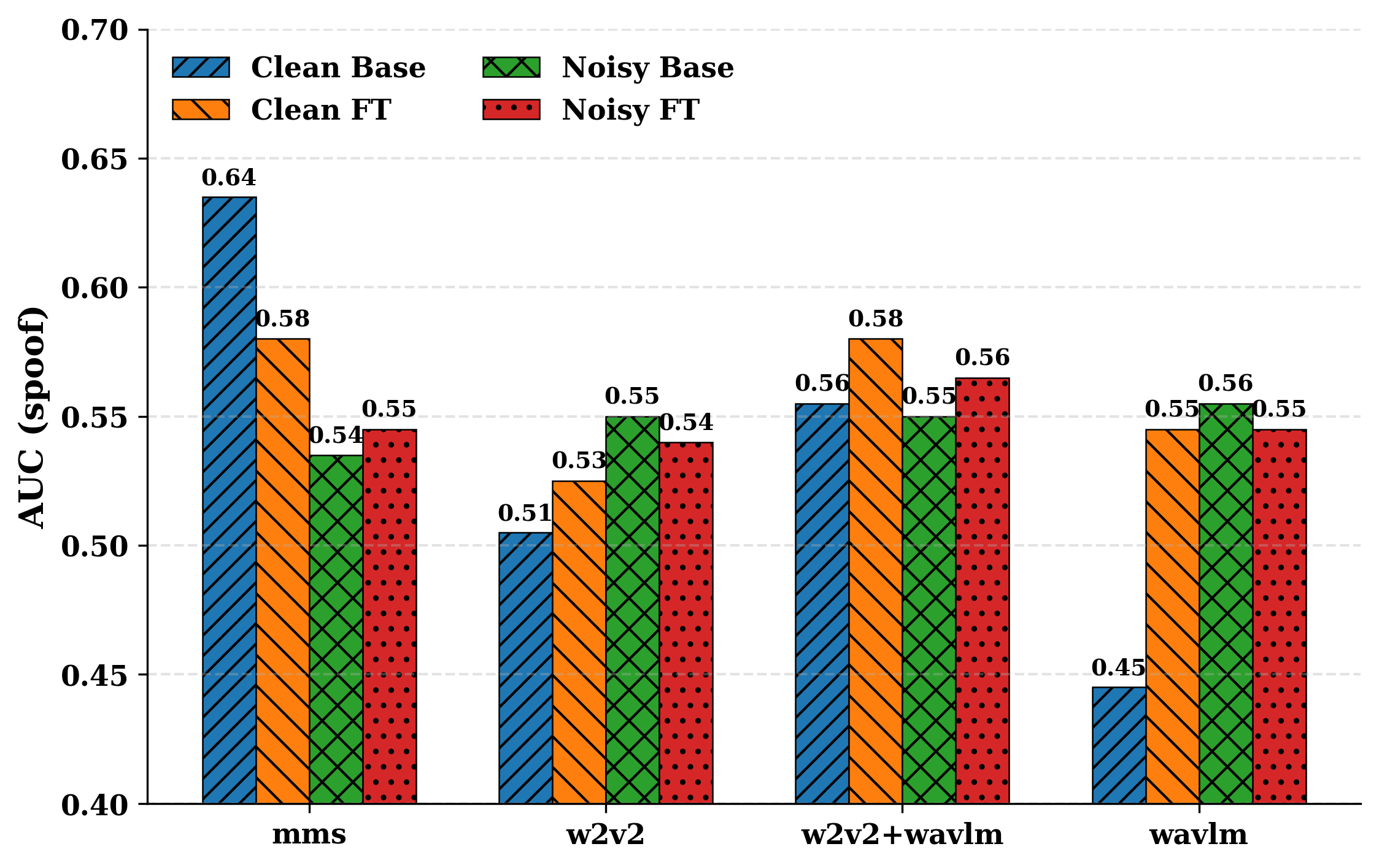}}    
  \caption{Binary (real vs.\ spoof) ROC--AUC under clean and noisy test conditions. Higher is better.}
  \label{fig:binary_auc}
\end{figure}

\subsection{Binary Detection (Clean and Noisy)}
On the clean-only test set, all encoders achieve strong baselines (low EER, high AUC). When evaluated on noise-augmented test data, frozen models degrade substantially at low SNRs (e.g., $\leq 0$\,dB), while finetuning consistently recovers performance. Across encoders, we observe a typical absolute EER reduction of $\sim$10--15\% at low SNR after finetuning. WavLM is the most robust among frozen baselines and continues to lead after finetuning. Figure~\ref{fig:binary_auc} compares clean vs.\ noisy ROC--AUC for the binary task across all backbones.

\subsubsection*{Per-SNR Spoof Robustness}
\begin{figure}[t]
  \centering
  \includegraphics[width=0.90\linewidth, height = 6 cm]{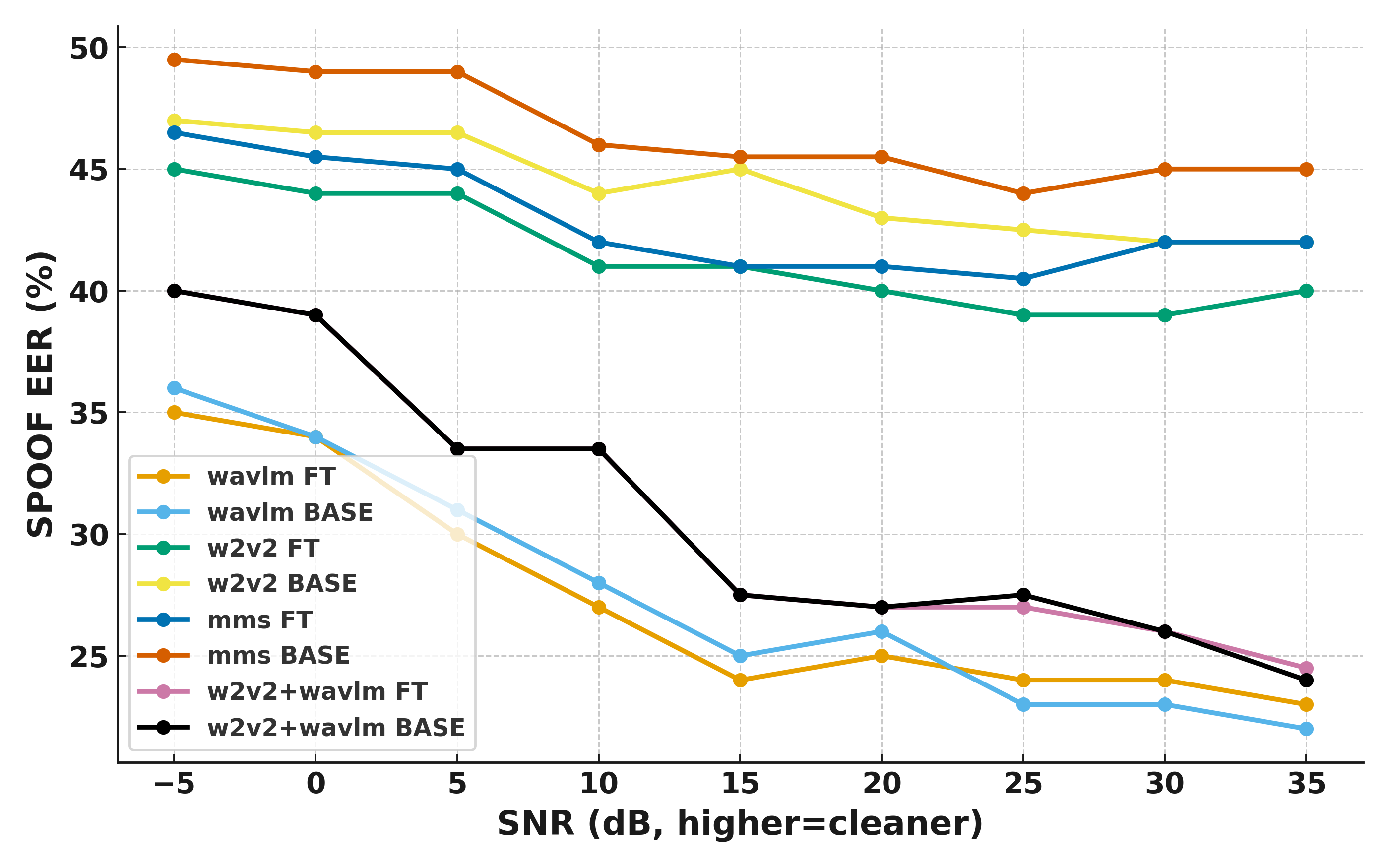}
  \caption{EER vs.\ SNR for \emph{spoof detection}: finetuned (solid) vs.\ baseline (dashed). Lower is better.}
  \label{fig:eer_snr_spoof}
\end{figure}

To isolate the effect of noise severity, we evaluate models at fixed SNRs. Figure~\ref{fig:eer_snr_spoof} plots EER as a function of SNR: finetuning consistently lowers and flattens the EER--vs--SNR curve, with the largest absolute gains between 10\,dB and 0\,dB, indicating improved robustness as SNR decreases.

% ===================== EER vs SNR (spoof) — Fig. 5 =====================

\subsection{\textbf{Four-class Detection (Authenticity $\times$ Corruption)}}
Frozen encoders tend to confuse Real+Noisy with Spoof+Noisy, indicating limited noise invariance. Finetuning improves both axes: (i) spoof detection within each noise condition and (ii) noise-state identification within each authenticity class. We see 15--20 point gains in macro-F1 on the four-class task on average (exact gains depend on the backbone and SNR). A two-layer MLP head further improved separability for some backbones, suggesting mild head capacity helps when authenticity and corruption cues interact.

% ===================== EER vs SNR (noise) — Fig. 6 =====================
\begin{figure}[htbp]  
  \centering
  \includegraphics[width=0.90\linewidth, height = 6 cm]{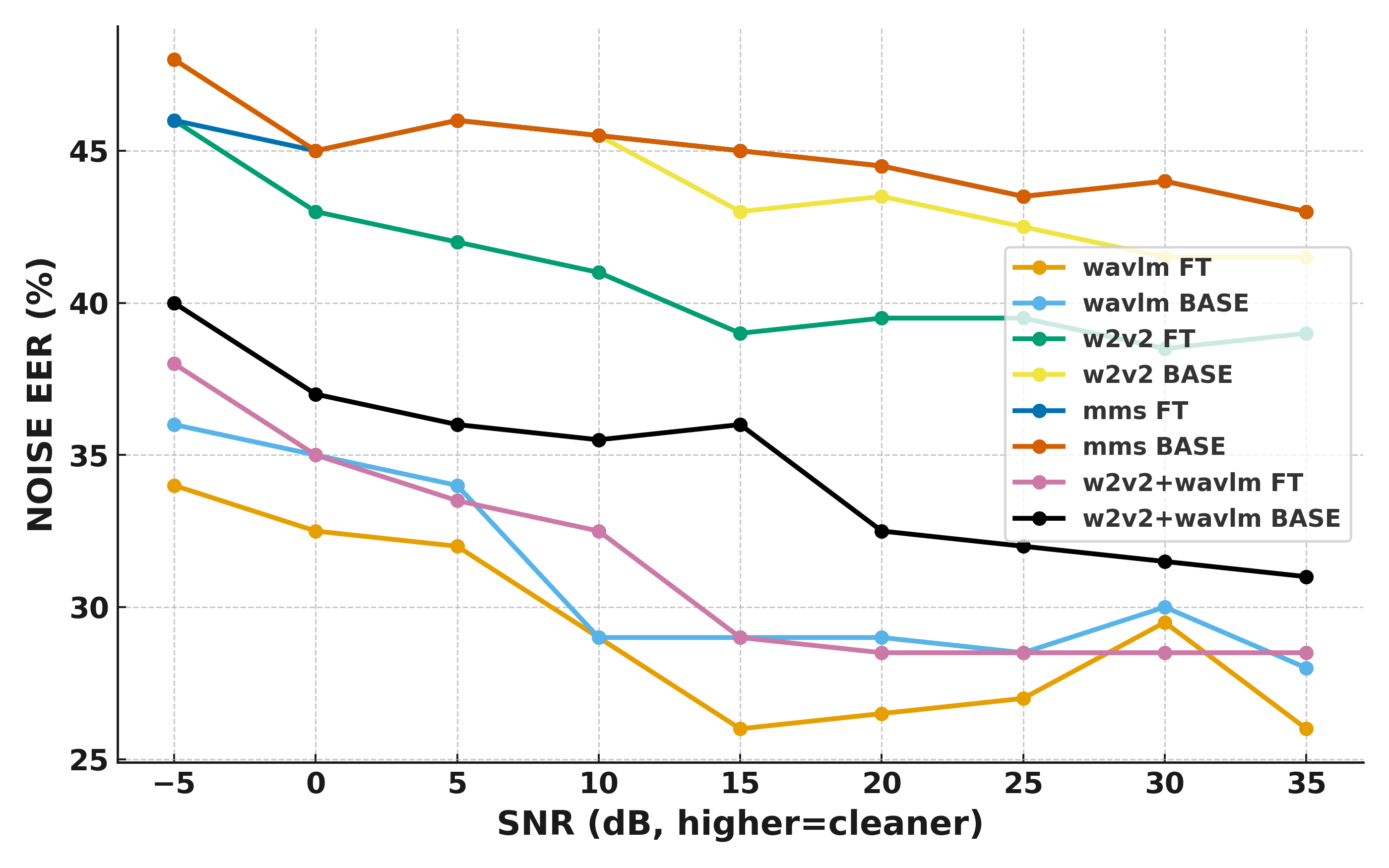}
  \caption{EER vs.\ SNR for \emph{noise discrimination} (clean vs.\ noisy): finetuned (solid) vs.\ baseline (dashed). Lower is better.}
  \label{fig:eer_snr_noise}
\end{figure}

Figure~\ref{fig:eer_snr_noise} shows that, across fixed SNR conditions, finetuning consistently lowers EER for the noise discrimination task compared with the frozen baseline, indicating stronger robustness at all tested SNRs.

\subsection{Ablations and Hyperparameters}
Two-stage LR (head-only $\rightarrow$ end-to-end) is consistently better than a single LR throughout. Balanced sampling reduces label skew effects, especially for the four-class setting. We found modest benefits from (i) slightly longer segments (4\,s vs.\ 2\,s) and (ii) mixing two noise clips occasionally, both of which increase invariance to nonstationary noise.

\subsection{Error Analysis}
To better understand model behavior, we examined confusion matrices and error patterns. A key finding is that \textbf{Real+Noisy utterances are often misclassified as Spoof+Noisy}, especially for frozen encoders. This indicates that background noise introduces distortions resembling spoof artefacts, reducing authenticity separability. Misclassifications are also concentrated at \textbf{low SNR levels ($\leq$\,5\,dB)}. At these conditions, background noise can dominate spectral regions where spoof artefacts (e.g., vocoder phase errors, high-frequency irregularities) are normally detected. As a result, the discriminative cues exploited by detectors are masked, leading to elevated error rates. Explicit four-class supervision mitigates this effect but does not fully eliminate the problem, pointing to a need for noise-aware feature extraction.

% ===================== p_noisy sensitivity (spoof) — Fig. 7 =====================
\begin{figure}[hbt]
  \centering
  \includegraphics[width=\columnwidth]{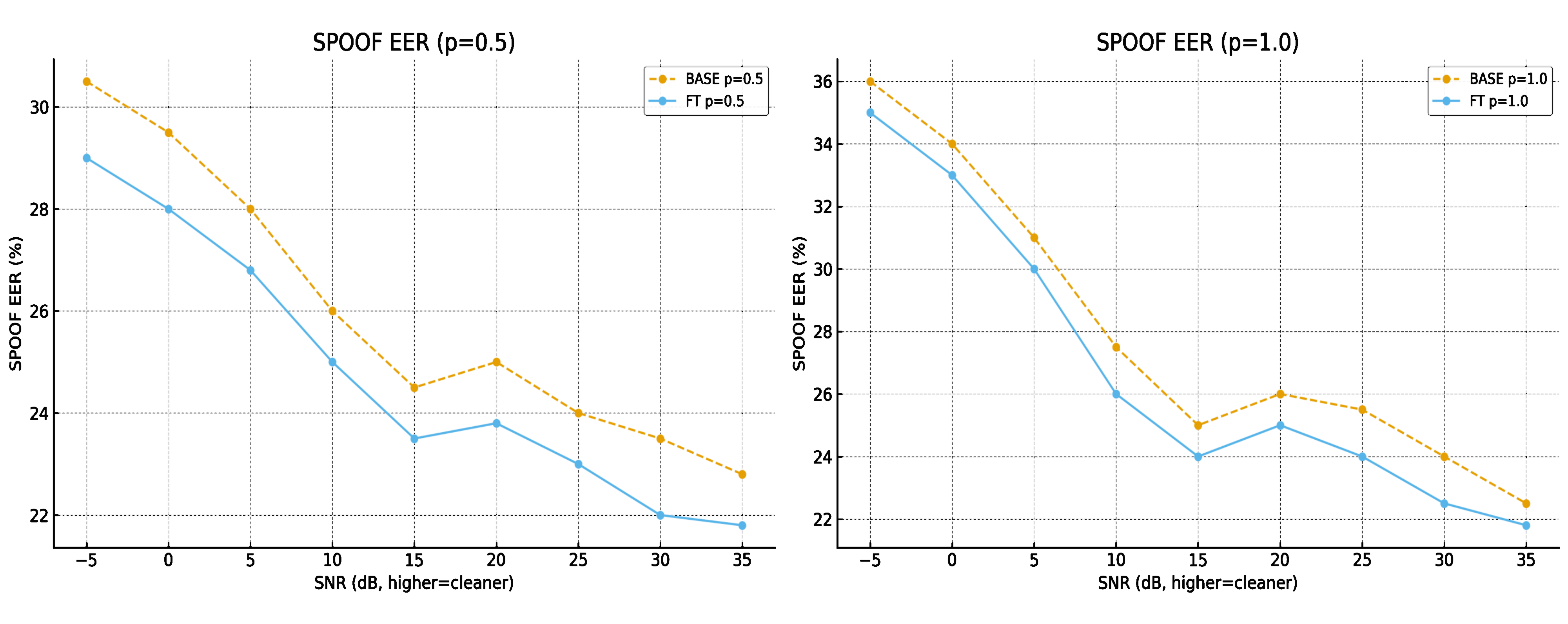}
  \caption{Effect of the noisy-trial proportion $p_{\text{noisy}}$ on \emph{spoof} performance (example backbone: WavLM). Lower curves indicate better robustness.}
  \label{fig:pnoisy_spoof}
\end{figure}

Figure~\ref{fig:pnoisy_spoof} summarizes spoof EER sensitivity to $p_{\text{noisy}}$, with mid-range values avoiding overfitting to either regime.

% ===================== p_noisy sensitivity (noise) — Fig. 8 =====================
\begin{figure}[hbt]
  \centering
  \includegraphics[width=\columnwidth]{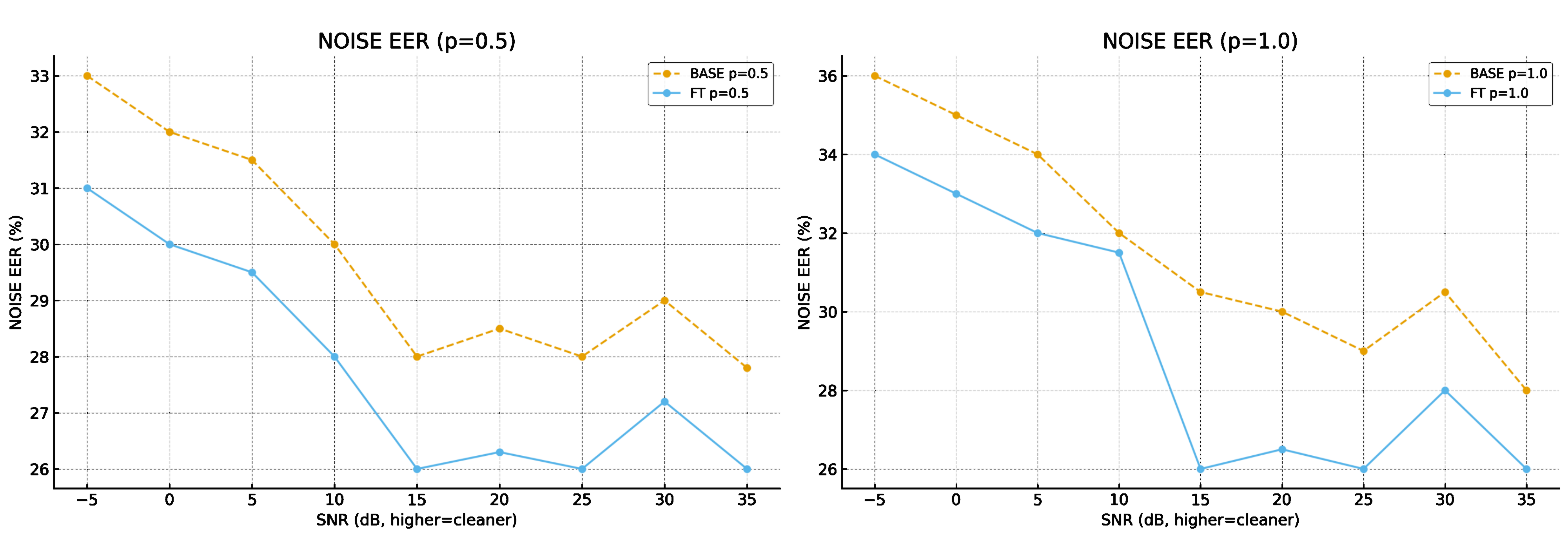}
  \caption{Sensitivity to the noisy-trial proportion $p_{\text{noisy}}$ for \emph{noise discrimination}. Lower is better.}
  \label{fig:pnoisy_noise}
\end{figure}

Figure~\ref{fig:pnoisy_noise} reports noise discrimination EER versus $p_{\text{noisy}}$, where balanced mixes yield the best trade-offs.

% ===================== Discussion and Future Work =====================
% ===================== Discussion =====================
\section{Discussion}

\noindent\textbf{Frozen SSL encoders are not inherently noise-robust.}
When encoders are used with a frozen configuration, the performance degrades sharply at lower SNR's (e.g., $\leq 5$\,dB). Although pretrained features capture prosodic and phonetic structures well, they are not guaranteed to encode noise-invariant spoof related cues. This highlights the need for explicit corruption aware training. 

\noindent\textbf{Finetuning substantially improves low-SNR behavior.}
End-to-end finetuning has shown consistently to stabilize the model under moderate and heavy noise. The largest gains typically emerge in the mid- and low-SNR ranges (10\,dB to 0\,dB), where the encoder becomes more selective to spoof-related structure while discounting stationary noise. 

\noindent\textbf{Increasing head capacity aids disentanglement.}
A single linear head may underfit when corruption and authenticity cues interact. However, introducing a lightweight two-layer MLP head improves separability in the four-class setup and provides more stable per-SNR performances, which indicates that mild non-linear transformations help isolate the underlying factors. 

\noindent\textbf{Four-class supervision improves discrimination under corruption.}
Explicitly labelling both authenticity and cleanliness  (Real+Clean, Real+Noisy, Spoof+Clean, Spoof+Noisy) reduces confusion between noisy real and noisy spoof conditions. This structured supervision benefits both the authenticity task and the related noise-estimation tasks.

\noindent\textbf{Noisy-mix probability acts as a controllable robustness parameter.}
The ratio of noisy to clean samples (\(p_{\text{noisy}}\)) determines the balance between clean accuracy and noise robustness. Moderate mixing proportions avoid regime-specific overfitting, and underscores the importance of reporting \(p_{\text{noisy}}\) in robustness-oriented evaluations.

% ===================== Future Work =====================

\section{Future Work}

\noindent\textbf{Multi-factor corruption robustness beyond additive noise.}
While this work focuses on additive environmental noise, real-world audio degradation involves a richer set of corruptions, including codec compression, reverberation, microphone mismatch, and even adversarial perturbations. Future evaluations should incorporate multi-factor stress tests that jointly vary codecs, room acoustics, channels, and device characteristics. Such comprehensive perturbation suites would better approximate deployment scenarios and enable detectors that remain reliable under complex, compounded distortions.

\noindent\textbf{Corruption-aware representation learning and adaptation.}
A promising direction is to integrate corruption-awareness directly into self-supervised learning objectives. Multi-corruption SSL pretraining—spanning noise, reverb, codec, and channel perturbations—may yield representations that preserve spoof artefacts even under severe distortion. Complementary strategies such as feature-space alignment, domain-mixup, style randomization, and simulated environments (e.g., room impulse responses) can improve channel and device invariance. Multi-view consistency losses inspired by CPC or SimCLR may further encourage stable embeddings across corrupted views of the same utterance.

\noindent\textbf{Toward standardized, multimodal robustness benchmarks.}
The field lacks unified protocols for evaluating corruption robustness in audio deepfake detection. A standardized benchmark combining noisy speech corpora (e.g., MUSAN, DEMAND), codec variants, telephony channels, and multi-condition test splits would support fair and reproducible comparison across models. Extending such benchmarks to audiovisual settings—studying how noise interacts with lip–speech synchronization or cross-modal consistency—remains an underexplored but essential direction for building deployable deepfake detection systems.

\section{Conclusion}
We presented a survey and a reproducible SNR-controlled benchmark for audio deepfake detection. Across WavLM, Wav2Vec2, and MMS encoders, \emph{frozen} baselines are brittle under noise, while \emph{finetuning} with on-the-fly augmentation substantially improves robustness, especially at 10--0\,dB SNR. Four-class supervision aids disentanglement of authenticity and corruption, and small increases in head capacity (two-layer MLP) further help where cues are entangled. The framework provides a practical recipe and a basis for future noise-aware pretraining and signal-processing–augmented systems.

% \section*{Acknowledgments}
% This work was supported by the \textit{Nanyang Technological University (NTU) Research Scholarship} and by \textit{Deepfaic}. We thank colleagues for insightful discussions.

\bibliographystyle{IEEEtran}
\bibliography{refs}

\end{document}